\documentclass[]{aa}
\usepackage{graphicx,natbib}

\def\jh{\mbox{$\rm (J-H)$}}

\def\jk{\mbox{$\rm (J-K_s)$}}

\def\ebv{\mbox{$\rm E(B-V)$}}
\def\ejh{\mbox{$\rm E(J-H)$}}
\def\rc{\mbox{$\rm R_{core}$}}
\def\rl{\mbox{$\rm R_{lim}$}}

\def\ms{\mbox{$\rm M_\odot$}}
\def\ds{\mbox{$\rm d_\odot$}}
\def\tdis{\mbox{$\rm t_{dis}$}}
\def\tr{\mbox{$\rm t_{rel}$}}
\def\Rgc{\mbox{$\rm R_\odot$}}
\def\dgc{\mbox{$\rm R_{GC}$}}

\def\jj{\mbox{$\rm J$}}
\def\hh{\mbox{$\rm H$}}
\def\ks{\mbox{$\rm K_s$}}

\begin{document}

\title{Old open clusters in the inner Galaxy: FSR\,1744, FSR\,89 and FSR\,31}

\author{C. Bonatto\inst{1} \and E. Bica\inst{1}}
\offprints{Ch. Bonatto}

\institute{Universidade Federal do Rio Grande do Sul, Departamento de Astronomia\\
CP\,15051, RS, Porto Alegre 91501-970, Brazil\\
\email{charles@if.ufrgs.br, bica@if.ufrgs.br}
\mail{charles@if.ufrgs.br} }

\date{Received --; accepted --}

\abstract
{The dynamical survival of intermediate-age/old open clusters in the inner Galaxy.}
{To establish the nature and derive fundamental and structural parameters of the
recently catalogued objects FSR\,1744, FSR\,89 and FSR\,31. This work intends to 
provide clues to constrain the Galactic tidal disruption efficiency, improve statistics 
of the open cluster parameter space, and better define their age-distribution function
inside the Solar circle. The current status of the issue dealing with the small number 
of detected open clusters in the inner Galaxy is discussed.}
{Properties of the objects are investigated by means of 2MASS colour-magnitude diagrams
and stellar radial density profiles built with field star decontaminated photometry.
Diagnostic-diagrams for structural parameters are used to help disentangle dynamical
from high-background effects affecting such centrally projected open clusters.}
{FSR\,1744, FSR\,89 and FSR\,31 are Gyr-class open clusters located at Galactocentric
distances $4.0 - 5.6$\,kpc. Compared to nearby open clusters, they have small core and
limiting radii.}
{With respect to the small number of open clusters observed in the inner Galaxy, the 
emerging scenario in the near-infrared favours disruption driven by dynamical evolution 
rather than observational limitations associated with absorption and/or high background 
levels. Internally, the main processes associated with the dynamical evolution are, e.g. 
mass loss by stellar evolution, mass segregation and evaporation. Externally they are, e.g. 
tidal stress from the disk and bulge, and interactions with giant molecular clouds. FSR\,1744, 
FSR\,89 and FSR\,31 have structural parameters consistent with their Galactocentric distances, 
in the sense that tidally induced effects may have accelerated the dynamical evolution.}


\titlerunning{Old open clusters in the inner Galaxy}

\authorrunning{C. Bonatto and E. Bica}

\maketitle

\section{Introduction}
\label{intro}

Star clusters evolve dynamically because of internal (e.g. mass loss during stellar 
evolution, mass segregation and evaporation) and external (e.g. tidal interactions with the disk and 
Galactic bulge, and collisions with giant molecular clouds - GMCs) processes. As clusters age, their 
structure suffer significant changes, to the point that most end up completely dissolved in the Galactic 
stellar field or as poorly-populated remnants.

Probably reflecting the Galactocentric distance-dependent strength of the external tidal stress, there 
is an asymmetry in the age distribution of open clusters (OCs) in the Galaxy: old ($\rm age>1$\,Gyr)
OCs are scarce and tend to be more concentrated towards the Galactic anti-center than young ones 
(e.g. \citealt{vdBMc80}; \citealt{Friel95}; \citealt{DiskProp}).

Historically, disruption was the primary effect assumed to explain the scarcity of OCs older than
$\sim10^8 - 10^9$\,yr in the Galaxy, a fact that was almost simultaneously noted by \citet{vdB57},
\citet{Oort58} and \citet{vHoerner58}. On theoretical grounds, \citet{Spitzer58} calculated disruption
time-scales (\tdis) consistent with the old-age cutoff and indicated that collisions with interstellar
clouds should also contribute to cluster disruption. Disruptive effects due to GMCs were investigated
by \citet{Wielen71} and \citet {Wielen91}, who estimated $\tdis\sim200$\,Myr for $50\%$ of the Milky Way
OCs, and suggested a dependence of \tdis\ on cluster size and mass. \citet{Gieles06} found $\tdis\sim2$\,Gyr
for the disruption of a $10^4\,\ms$ cluster due to encounters with GMCs, and suggested that the combined
effect of tidal field and encounters with GMCs can explain the lack of old OCs in the solar neighbourhood.
Further observational evidence for the dependence of the disruption time-scale on mass are given by
\citet {JA82} and \citet{Lamers05} (and references therein). The latter authors found $\tdis\sim M^{0.62}$, 
which for cluster mass in the range $10^2 - 10^3\ms$ corresponds to $\rm75\la\tdis(Myr)\la300$.

Indeed, the well-studied clusters in the central Galaxy are as a rule young and massive, which have 
not yet significantly responded to tidal effects. The Arches cluster, near the Galactic nucleus, has 
an age of 2\,Myr and a mass $M=10^5\,\ms$ (\citealt{Figer02}), while the Quintuplet cluster has 4\,Myr 
and an estimated mass $M=1.3\times10^4\,\ms$ (\citealt{Figer99}). BD\,11 is not as massive as the 
Arches cluster, but is similarly young and close to the center (\citealt{DOBBZM03}).

The disk ring $\rm1 < \dgc(kpc) < 3$ contains the noteworthy clusters Westerlund\,1 and BDSB\,122. 
Westerlund\,1 is located at $\ds\approx5$\,kpc with an age of $4.5-5$\,Myr (\citealt{Crowther06}). Its
initial mass is estimated to be $M\approx10^5\,\ms$ (\citealt{Clark05}). \citet{REF010} discovered the
cluster BDSB\,122 and \citet{Figer06} estimated it to be very massive with $M\approx2-4\times10^4\,\ms$.
BDSB\,122 contains 14 red supergiants with an age of $7-12$\,Myr at $\ds=5.8$\,kpc.

Despite the above observational and theoretical efforts, old OCs within $\la4$\,kpc of the Galactic centre
remain virtually undetected (e.g. \citealt{DiskProp}). Although conceptually different processes, tidal
shocks (from the Galaxy and GMCs) and observational completeness may combine to explain the scarcity 
of central clusters. Absorption and crowding in fields dominated by the disk and bulge stars are expected to
significantly decrease completeness, especially at the faint-end of the OC luminosity distribution.
Indeed, using a sample of 654 OCs with available fundamental parameters, \citet{DiskProp} found that a 
large fraction of the intrinsically faint and/or distant OCs must be drowned in the field contamination,
particularly in bulge/disk directions.

Undoubtedly, the nuclear clusters are doomed to dissolution (\citealt{Portegies02}). In fact, based on
arguments related to observational completeness affecting the spatial distribution of the Galactic OCs, 
\citet{DiskProp} found that tidal disruption may begin to be important for OCs located at distances 
$\ga1.4$\,kpc inside the Solar circle. A fundamental question is whether massive clusters in the 
intermediate ring $\rm1 < \dgc(kpc) < 3$, like presently Westerlund\,1 and BDSB\,122, might survive to 
old ages.

Recently, \citet{FSR07} have presented a catalogue of 1021 cluster candidates (hereafter FSR objects)
at Galactic latitudes $|b|<20^\circ$ and all longitudes, by means of an automated algorithm applied to the
2MASS\footnote{The Two Micron All Sky Survey, available at {\em www.ipac.caltech.edu/2mass/releases/allsky/ }}
database. In the present study we search for old OC candidates towards the central parts of the Galaxy from
the targets in that catalogue. Detection of such objects will help constrain the dynamical survival issue.

Recent determinations of the distance of the Sun to the Galactic center by means of different methods
are decreasing as compared to the usual $\Rgc=8.0$\,kpc (\citealt{Reid93}). \citet{Eisenhauer05} found
$\Rgc=7.6\pm0.3$\,kpc, \citet{Nishiyama06} obtained $\Rgc=7.5\pm0.35$\,kpc, and \citet{GCProp}
$\Rgc=7.2\pm0.3$\,kpc. The latter value, based on updated distances of globular clusters, is adopted
throughout the present paper.

\begin{figure}
\resizebox{\hsize}{!}{\includegraphics{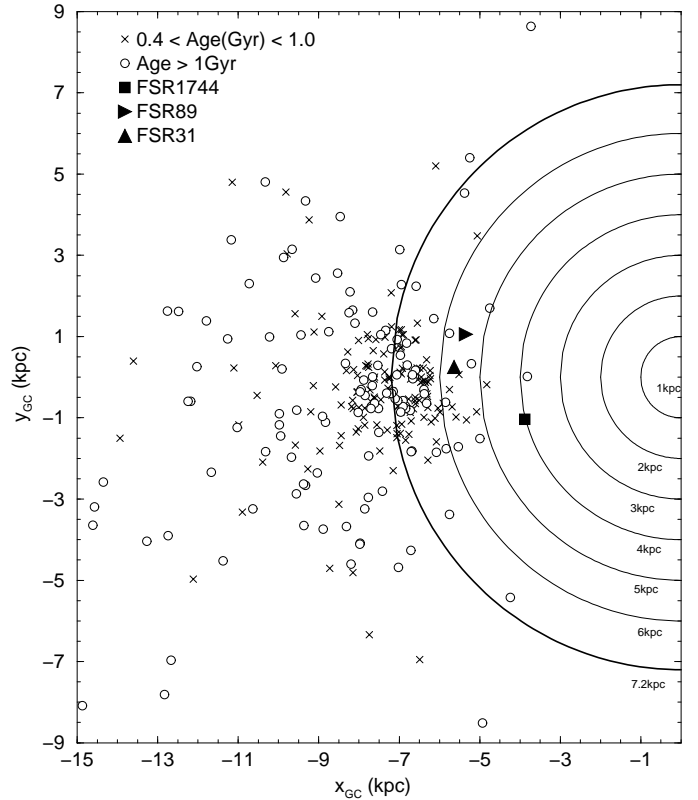}}
\caption{Spatial distribution of the WEBDA OCs with ages $0.4 - 1$\,Gyr (crosses) and older than 1\,Gyr
(empty circles). The semi-circles indicate Galactocentric distances from 1 to 7.2\,kpc. The position
of the candidates to old OCs FSR\,1744, FSR\,89 and FSR\,31 are shown (Sect.~\ref{PhotPar}).}
\label{fig1}
\end{figure}

\begin{figure}
\resizebox{\hsize}{!}{\includegraphics{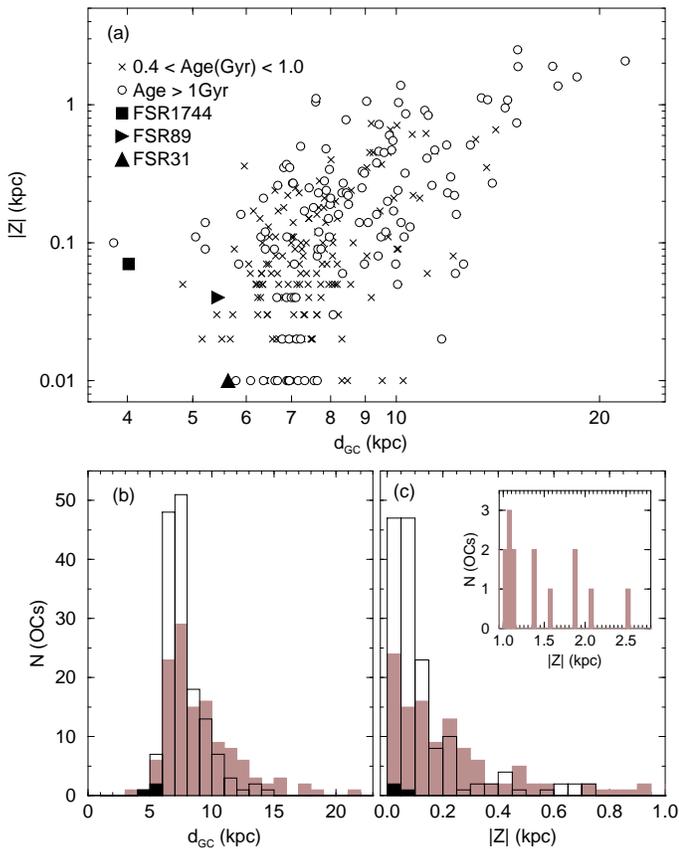}}
\caption{Panel (a): Relation of distance from the plane $|Z|$ with \dgc\ for the OCs older than 1\,Gyr 
(empty circles) and those with age in the range $0.4-1.0$\,Gyr (crosses). The respective \dgc\ and $|Z|$ 
histograms are shown in panels (b) and (c) for OCs older than 1\,Gyr (gray histograms), those with ages 
$0.4-1.0$\,Gyr (empty), and the present objects (black). The inset in panel (c) contains the OCs at 
$|Z|>1.0$\,kpc. These are all older than 1\,Gyr.}
\label{fig2}
\end{figure}

In the present work we investigate the nature of three candidates to old OCs in the inner Galaxy by means 
of near-infrared colour-magnitude diagrams (CMDs) and cluster structural parameters. In Sect.~\ref{DynEvol}
we discuss aspects related to the dynamical evolution and disruption of star clusters. In Sect.~\ref{OpticalOv} 
we discuss the spatial distribution of the known old OCs. In Sect.~\ref{PhotPar} we analyse CMDs and compute
cluster fundamental parameters. In Sect.~\ref{Struc} we derive structural parameters. Discussions on 
the nature of the objects are provided in Sect.~\ref{discuss}. Concluding remarks are given in 
Sect.~\ref{Conclu}.

\section{Considerations on dynamical evolution and disruption}
\label{DynEvol}

Star clusters form in collapsing molecular clouds in which variable fractions ($10-30\%$, e.g.
\citealt{LL2003}) of the parent gas are converted into stars. They remain embedded in the clouds 
for about $2-5$\,Myr, and their dynamical state at that early phase can be described as out of virial
equilibrium (e.g. \citealt{delaF2002}). Following the rapid expulsion of the unused gas by massive
winds and supernovae, stars end up with exceeding velocities with respect to the new, decreased,
potential (e.g. \citealt{delaF2002}; \citealt{BK2002}; \citealt{GoBa06}). As a consequence, clusters 
expand in all scales as they reach for virialization. Reflecting the violent structural changes
associated with this early phase, the number of optically detected clusters, i.e. those that survived 
the infant mortality, is significantly smaller than the number of embedded clusters ($\approx4\%$, 
\citealt{LL2003}).

N-body simulations of massive star clusters that include the effect of gas removal (e.g. 
\citealt{GoBa06}) consistently show that the phase of dramatic core radii increase may last about 10-30\,Myr. 
They also suggest that after a few $10^7$\,yr, core growth levels off as some energy equipartition is reached. 
The outer parts, on the other hand, keep expanding and become loosely bound. Theoretically,
the subsequent core evolution depends on several factors such as cluster mass and the relative efficiencies 
of energy equipartition, evaporation and gravothermal instability. These processes tend to produce
a phase of core contraction that may be followed by a rapid expansion, which is confirmed by N-body
simulations of clusters with less than 20\,000 particles and stars with two different masses (e.g.
\citealt{Khalisi07}). An alternative effect that also leads to core contraction, at least in the early
phases, is the violent relaxation that follows from the time dependent potential resulting from rapid gas
removal (\citealt{BasGod06}; \citealt{KrAaHur01}). The bottom line is that mass loss (from gas removal 
or stellar evolution) leads to cluster expansion in all scales at the early phases, while dynamical friction 
in a multi-mass system leads to core collapse/contraction at later phases.
 
With respect to other galaxies, most LMC and SMC clusters follow a trend of increasing core radius 
with cluster age, but there is as well observational evidence for clusters in a phase of core shrinkage 
occurring at several hundred Myr (e.g. \citealt{MG03}). Considering the very different time-scales 
involved, the early gas removal (\citealt{GoBa06}) cannot explain the core shrinkage at such later phases.

\begin{figure}
\resizebox{\hsize}{!}{\includegraphics{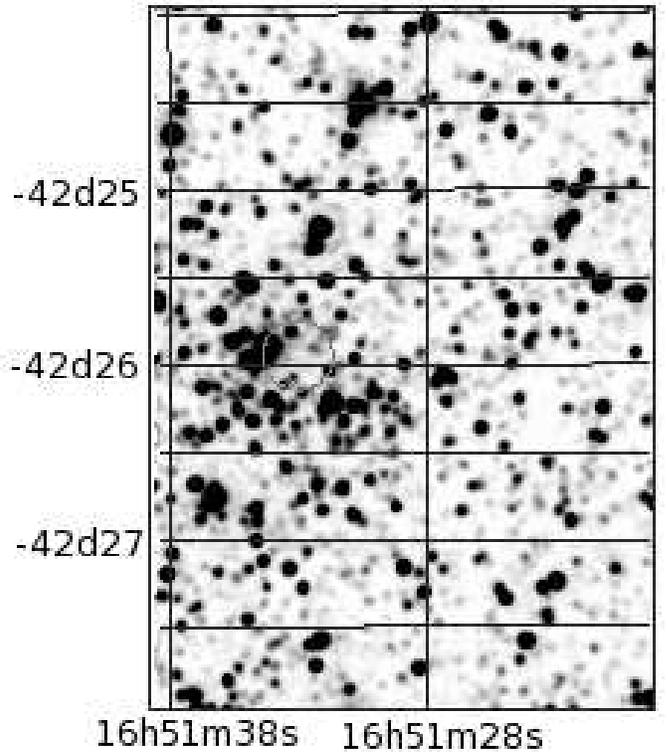}}
\caption{$4\arcmin\times4\arcmin$ 2MASS H image of FSR\,1744. Images provided by the 2MASS Image Service. 
The small circle indicates the optimized central coordinates (cols.~5 and 6 of Table~\ref{tab1}).}
\label{fig3}
\end{figure}

\begin{figure}
\resizebox{\hsize}{!}{\includegraphics{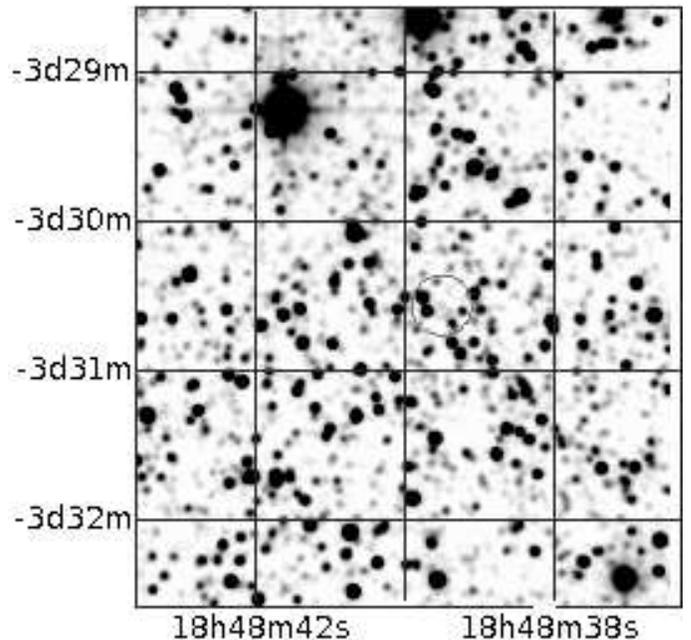}}
\caption{Same as Fig.~\ref{fig3} for FSR\,89.}
\label{fig4}
\end{figure}

\begin{figure}
\resizebox{\hsize}{!}{\includegraphics{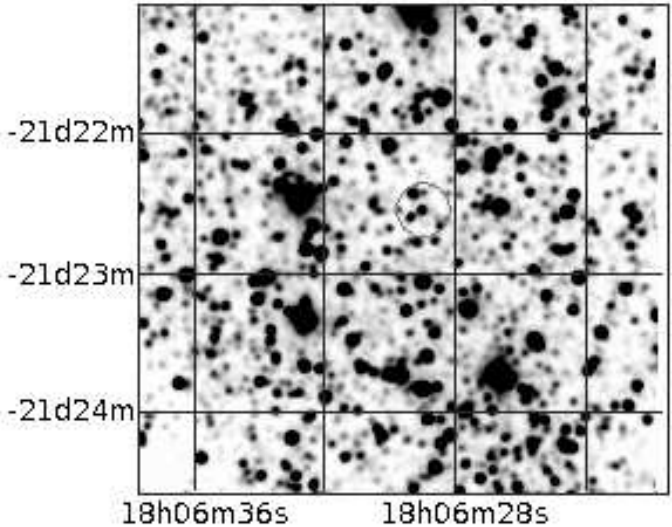}}
\caption{Same as Fig.~\ref{fig3} for FSR\,31.}
\label{fig5}
\end{figure}

Observational evidence of a dependence of the core mass function slope (and radius) on the dynamical-evolution 
parameter $\tau={\rm Age/\tr}$, where $\tr$ is the relaxation time, was found for a set of nearby OCs by
\citet{DetAnalOCs} and \citet{BB07}. The dependence occurs in the sense that dynamically old clusters tend to 
have shallow core mass functions and, for part of the clusters, small core radius. Similar relations were
found in a sample of 42 OCs by means of BV CCD photometry (\citealt{MacNie07}), and in 9 others with wide-field 
CCD photometry (\citealt{Sharma06}). 

Eventually, most clusters end up destroyed by internal processes (mass loss caused by the dynamical and 
stellar evolutions) or leave remnants (\citealt{PB07} and references therein). On the other hand, higher
efficiencies of star formation ($\ga30\%$) are predicted to enhance cluster survival rates. Such clusters 
can reach old ages, although the final clusters retain only a small fraction of the initial mass 
(\citealt{GoBa06}).

At the same time, interactions with the disk, the tidal pull of the Galactic bulge, and collisions
with GMCs (Sect.~\ref{intro}), tend to destroy the poorly-populated OCs in a time-scale of a few
$10^8$\,Myr (\citealt{Bergond2001}), especially in the inner Galaxy. Indeed, \citet{vdBMc80}
noted that OCs older than $\ga1$\,Gyr tended to be concentrated towards the anti-center, a region
with a low density of GMCs. The observation that the fraction of old clusters in the LMC and 
SMC is higher than in the Galaxy suggests that the disruption time-scale may not be the same in
all galaxies (\citealt{EF85}; \citealt{Wielen85}; \citealt{Hodge87}), instead, it may vary with 
the different GMC density in those galaxies.

Over large time periods, the net effect of tidal interactions on a star cluster is to heat it, in the
sense that its stars gain more kinetic energy after each event, which leads to an increase on the
evaporation rate. Recent works estimate that the central ($\dgc\la150$\,pc) tidal fields can dissolve 
a massive star cluster in $\approx50$\,Myr (\citealt{Portegies02}). Near the Solar circle most OCs 
appear to dissolve in a time-scale shorter than $\sim1$\,Gyr (\citealt{Bergond2001}; \citealt{DiskProp}), 
consistent with the $1.3\pm0.5$\,Gyr estimate for a $10^4\,\ms$ cluster found by \citet{Lamers05}. These 
time-scales are intermediate between those predicted from N-body simulations of OCs in the tidal field of 
the Solar vicinity ($\sim6.3$\,Gyr, \citealt{BM03}) and the $322\pm31$\,Myr derived for \citet{Piskunov06}. 
Differences among these time-scales probably reflect a dependence of disruption on cluster mass, since 
the relatively short time-scales of \citet{Piskunov06}, \citet{DiskProp} and \citet{Bergond2001} correspond 
to OCs of lower mass ($\la10^3\,\ms$) than the $10^4\,\ms$ used in the models of \citet{Lamers05}. Besides, 
the large time-scale of \citet{BM03} probably results because their models do not include external 
perturbations.

Except for the N-body time-scale, the remaining ones are consistent with the fact that old OCs
($\rm age\ga1$\,Gyr) are usually found about the Solar circle and in the outer Galaxy (e.g. \citealt{Friel95};
\citealt{DiskProp}), where the frequency of potentially damaging dynamical interactions with GMCs and
the disk is low (e.g. \citealt{Salaris04}; \citealt{Upgren72}). In addition, simulations show that the
initial mass function, fraction of binaries, total mass and Galactocentric distance are factors intimately
linked to cluster lifetimes (e.g. \citealt{Terlevich87}; \citealt{deLaF97}; \citealt{deLaF98}).
Considering these aspects, the dynamical survival rate of star clusters, especially in the inner
Galaxy, can possibly be used to probe the tidal field of the Galaxy.

\section{An overview of the evolved and old open clusters}
\label{OpticalOv}

The WEBDA\footnote{\em obswww.unige.ch/webda} database (\citealt{Merm1996}) provides updated parameters
for the studied optical OCs. We selected from WEBDA the OCs with age in the range $0.4 - 1$\,Gyr, and
those older than 1\,Gyr, for which we compute the projections on the Galactic plane of the Galactic
coordinates ($\ell,b)$ $x_{GC}=\ds\cos(b)\cos(\ell)-\Rgc$ and $y_{GC}=\ds\cos(b)\sin(\ell)$, where 
\ds\ is the distance from the Sun and $\Rgc=7.2$\,kpc (\citealt{GCProp}). The $x_{GC}$ and $y_{GC}$ 
distributions are shown in Fig.~\ref{fig1}. Both age groups share a similar spatial distribution, in 
which the number-density of OCs decreases with Galactocentric distance beyond the Solar circle (e.g.
\citealt{Friel95}; \citealt{DiskProp}).

A systematic inspection of the 2MASS images of the FSR candidates carried out by one of us (EB)
revealed that in the Galactic longitude zone $|l|<30^\circ$, the 3 objects FSR\,1744, FSR\,89 and FSR\,31,
showed up as probable OCs. Another interesting object in that zone is FSR\,1735, probably a globular
cluster (\citealt{FMS07}).

Figure~\ref{fig1} shows as well the positions of FSR\,1744, FSR\,89 and FSR\,31, based on the
Galactocentric distances and components computed in Sect.~\ref{PhotPar}. They are located at
$\rm4\la\dgc(kpc)\la6$, with FSR\,1744 being the innermost at $\dgc=4.0$\,kpc (Sect.~\ref{PhotPar}).

In Fig.~\ref{fig2} we further explore spatial properties of FSR\,1744, FSR\,89 and FSR\,31 with respect
to those of the OCs older than 0.4\,Gyr. The relation of the distance to the Galactic plane ($|Z|$) with
Galactocentric distance is shown in panel (a). Only old OCs are found at $\dgc\ga14$\,kpc and 
$|Z|\ga0.65$\,kpc, as previously noted by e.g. \citet{vdBMc80}. This should be expected, since OCs in the
outer disk and/or in regions high above the plane spend most of their orbits far away from the disruptive 
effects associated with the central Galactic regions and GMCs (e.g. \citealt{Friel95}). Panels (b) and (c)
contain histograms for the number of OCs (in each age range) as a function of Galactocentric distance and
distance to the Galactic plane, respectively. We are dealing with 3 old OCs located among the innermost
ones in the Galaxy. As already suggested by their low Galactic latitudes (Table~\ref{tab1}) they are within
$\approx100$\,pc of the plane. Consequently, they are expected to be affected by important field-star
contamination.

\section{Photometric parameters with 2MASS}
\label{PhotPar}

The original coordinates of FSR\,1744 (\cite{FSR07}) are slightly shifted with respect to the
central concentration of stars as seen in 2MASS images (e.g. Fig.~\ref{fig3}). The revised values,
together with the original ones for the three objects are given in Table~\ref{tab1}. Table~\ref{tab1} 
also provides the following parameters derived in the present work, by columns, cluster age (9), 
reddening \ebv\ (10), distance from the Sun (11), and Galactocentric distance (12), derived in the
subsequent analyses.

\begin{table*}
\caption[]{Fundamental parameters}
\label{tab1}
\renewcommand{\tabcolsep}{1.75mm}
\renewcommand{\arraystretch}{1.5}
\begin{tabular}{lcccccccccccc}
\hline\hline

&\multicolumn{3}{c}{FSR2007}&\multicolumn{8}{c}{Present results derived from 2MASS data}\\
\cline{2-4}\cline{6-13}
Cluster&$\alpha(2000)$&$\delta(2000)$&D&&$\alpha(2000)$&$\delta(2000)$&$\ell$&$b$&Age&$\ebv$&\ds&\dgc\\
&(hms)&($^\circ\arcmin\arcsec$)&(\arcmin)&&(hms)&($^\circ\arcmin\arcsec$)&($^\circ$)&($^\circ$)&(Gyr)&&(kpc)&(kpc)\\
(1)&(2)&(3)&(4)&&(5)&(6)&(7)&(8)&(9)&(10)&(11)&(12)\\
\hline
FSR\,1744&16:51:36&$-$42:24:55&2.2&&16:51:32&$-$42:25:57&342.68&$+$1.18&$1.0\pm0.1$&$2.56\pm0.12$&$3.5\pm0.1$
      &$4.0\pm0.1$ \\
      
FSR\,89&18:48:39&$-$03:30:34&2&&18:48:39&$-$03:30:34&29.49&$-$0.98&$1.0\pm0.1$&$2.94\pm0.10$&$2.2\pm0.1$
      &$5.4\pm0.1$  \\

FSR\,31&18:06:29&$-$21:22:33&6&&18:06:29&$-$21:22:33&8.91&$-$0.27&$1.1\pm0.2$&$1.47\pm0.09$&$1.6\pm0.1$
      &$5.6\pm0.1$  \\
\hline
\end{tabular}
\begin{list}{Table Notes.}
\item Cols.~2 and 3: Central coordinates provided by \citet{FSR07}. Col.~4: angular diameter estimated on
the 2MASS images (Figs.~\ref{fig3} - \ref{fig5}). Cols.~5-8: Optimized central coordinates (from 2MASS 
data). Col.~10: reddening in the object's central region (Sect.~\ref{PhotPar}). Col.~12: \dgc\ calculated 
using $\Rgc=7.2$\,kpc (\citealt{GCProp}) as the distance of the Sun to the Galactic centre.
\end{list}
\end{table*}

2MASS H images covering a $4\arcmin\times4\arcmin$ field of the target objects are shown in
Figs.~\ref{fig3} - \ref{fig5}. In all cases a concentration of stars is superimposed on crowded 
fields, as expected from such central directions, especially for FSR\,31 (Table~\ref{tab1}).

\subsection{2MASS Photometry}
\label{2mass}

\jj, \hh\ and \ks\ 2MASS photometry was extracted in circular fields centred on the optimized
coordinates of the objects (cols.~5 and 6 of Table~\ref{tab1}) using VizieR\footnote{\em
vizier.u-strasbg.fr/viz-bin/VizieR?-source=II/246}. Our previous experience with OC analysis (e.g.
\citealt{BB07}, and references therein) shows that as long as no other populous cluster is present
in the field, and differential absorption is not prohibitive, wide extraction areas can provide the
required statistics, in terms of magnitude and colours, for a consistent field star decontamination
(Sect.~\ref{CMDs}). They also produce more constrained stellar radial density profiles (Sect.~\ref{Struc}).
The extraction radii are 20\arcmin, 20\arcmin, and 50\arcmin, respectively for FSR\,1744, FSR\,89 and 
FSR\,31. Such radii are $\approx30$ and $\approx4$ times larger than the respective core and limiting 
radii (Sect.~\ref{Struc}).

As photometric quality constraint, 2MASS extractions were restricted to stars with magnitudes {\em (i)}
brighter than those of the 99.9\% Point Source Catalogue completeness limit\footnote{Following the Level\,1
Requirement, according to {\em www.ipac.caltech.edu/2mass/releases/allsky/doc/sec6\_5a1.html }} in
the cluster direction, and {\em (ii)} with errors in \jj, \hh\ and \ks\ smaller than 0.2\,mag. The 99.9\%
completeness limits are different for each cluster, varying with Galactic coordinates.

\begin{figure}
\resizebox{\hsize}{!}{\includegraphics{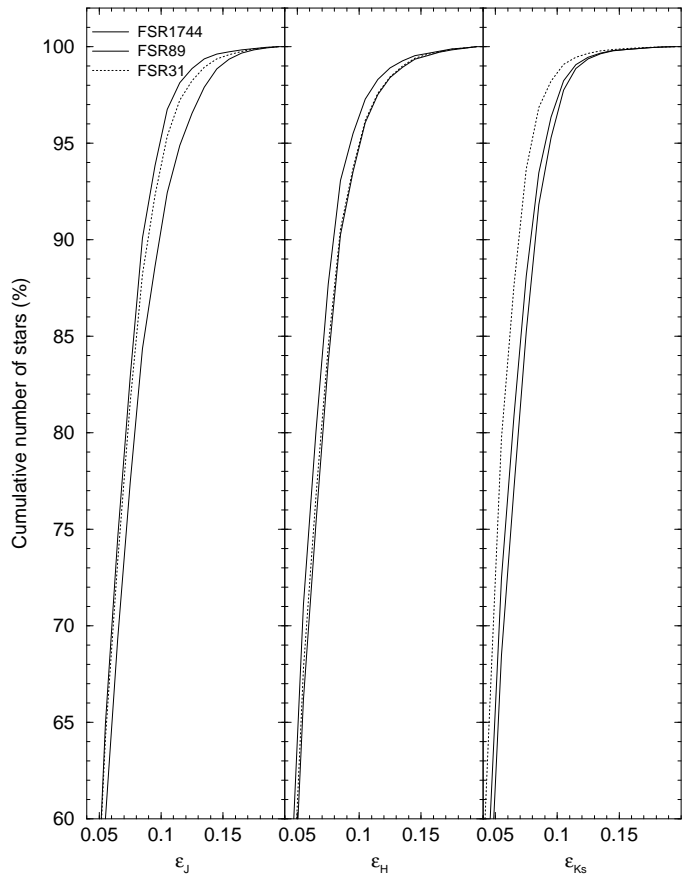}}
\caption{2MASS photometric errors evaluated by means of cumulative histograms with
the fraction of stars as a function of uncertainties. In all cases, a large fraction of the stars have 
uncertainties smaller than 0.06\,mag.}
\label{fig6}
\end{figure}

To objectively characterize the distribution of 2MASS photometric uncertainties in the fields of the
present objects, we show in Fig.~\ref{fig6} cumulative histograms with the fraction of stars as a 
function of uncertainties for the 3 bands. Projected areas sampled in the histograms correspond to 
the respective extraction radius of each object. The distributions of photometric uncertainties are
similar both among the fields sampled and 2MASS bands. The fraction of stars with \jj, \hh\ and \ks\ 
uncertainties smaller than $0.06$\,mag in FSR\,1744, FSR\,89 and FSR\,31 ranges from $\approx75\% - 85\%$.

For reddening transformations we use the relations $A_J/A_V=0.276$, $A_H/A_V=0.176$, $A_{K_S}/A_V=0.118$, 
and $A_J=2.76\times\ejh$ (\citealt{DSB2002}), assuming a constant total-to-selective absorption ratio $R_V=3.1$.

\subsection{CMD morphology and field-star decontamination}
\label{CMDs}

\subsubsection{FSR\,1744}
\label{FSR1744}

\begin{figure}
\resizebox{\hsize}{!}{\includegraphics{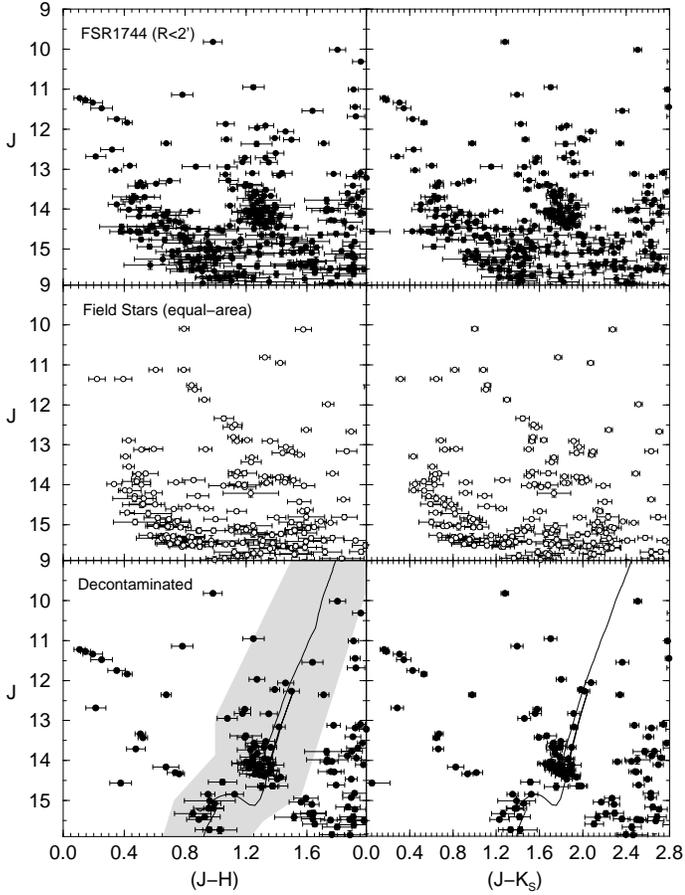}}
\caption{2MASS CMDs extracted from the $R<2\arcmin$ region of FSR\,1744. Top panels: observed photometry
with the colours $\jj\times\jh$ (left) and $\jj\times\jk$ (right). Middle: equal-area comparison field. Besides
contamination of disk and bulge stars, evidence of a giant clump shows up at $1.1\protect\la\jh\protect\la1.5$,
$1.5\protect\la\jk\protect\la2.0$ and $13.3\protect\la\jj\protect\la14.7$. Bottom panels: field star decontaminated
CMDs set with the 1\,Gyr Padova isochrone (solid line), where the giant clump and the top fraction of the MS are
conspicuous in both colours. The colour-magnitude filter used to isolate cluster MS/evolved stars is shown as a
shaded region.}
\label{fig7}
\end{figure}

2MASS $\jj\times\jh$ and $\jj\times\jk$ CMDs extracted from a central ($R<2\arcmin$) region of FSR\,1744
are presented in Fig.~\ref{fig7}. In this extraction that contains the bulk of the cluster
stars (Sect.~\ref{Struc}), a disk-like population ($0.2\la\jh\la0.8$ and $\jj\la15$) appears to merge into 
a redder component ($\jh\ga0.8$). With respect to the equal-area comparison field extraction, the red component
clearly presents an excess in the number of stars for $1.1\la\jh\la1.5$ and $13.3\la\jj\la14.7$ (top-left
panel), which resembles a giant clump of an intermediate-age OC. Similar features are present in
the $\jj\times\jk$ CMD (top-right panel).

To uncover the intrinsic CMD morphology we apply the field star decontamination algorithm described in
\cite{BB07}. The algorithm works on a statistical basis that takes into account the relative number-densities
of probable field and cluster stars in small cubic CMD cells. Cell axes correspond to the magnitude \jj\ and the
colours \jh\ and \jk\ (considering as well the $1\sigma$ uncertainties in the 2MASS bands). These are the
2MASS colours that provide the maximum variance among CMD sequences for OCs of different ages (e.g.
\citealt{TheoretIsoc}).

Basically, the algorithm {\em (i)} divides the full range of magnitude and colours of a given CMD into a 3D
grid, {\em (ii)} computes the expected
number-density of field stars in each cell based on the number of comparison field stars with magnitude and
colours compatible with those of the cell, and {\em (iii)} randomly subtracts the expected number of field
stars from each cell. The algorithm is sensitive to local variations of field-star contamination with colour
and magnitude (\citealt{BB07}). Typical cell dimensions are $\Delta\jj=0.5$, and $\Delta\jh=\Delta\jk=0.2$.
These values are large enough to allow sufficient star-count statistics in individual cells and small enough
to preserve the morphology of different CMD evolutionary sequences. Besides, 2MASS photometric uncertainties
for most stars of the present objects are considerably smaller than the adopted cell dimensions (Fig.~\ref{fig6}).
As comparison field we use the region $6<R(\arcmin)<20$ around the cluster center to obtain representative
background star-count statistics.

To minimize potential artifacts introduced by the choice of parameters, the algorithm is applied
for 3 different grid specifications in each dimension. For instance, for a CMD grid starting at magnitude
$J_o$ (and cell width $\Delta\jj$), additional runs for $J_o\pm\frac{1}{3}\Delta\jj$ are included. Considering
as well the 2 colours, 27 different outputs are obtained, from which the average number of probable cluster stars
$\langle N_{cl}\rangle$ results. Typical standard deviations of $\langle N_{cl}\rangle$ are at the $\approx2.5\%$
level. The final field star-decontaminated CMD contains the $\langle N_{cl}\rangle$ stars with the highest
number-frequencies. Stars that remain in the CMD after the field star decontamination are in cells
where the stellar density presents a clear excess over the field. Consequently, they have a significant probability
of being cluster members. Further details on the algorithm, including discussions on subtraction efficiency and
limitations, are given in \citet{BB07}.

The resulting field star decontaminated CMDs are shown in the bottom panels of Fig.~\ref{fig7}. As expected,
most of the disk component is removed, leaving two groups of stars that resemble the top of a main sequence
(MS) and a giant clump. Residual contamination, mostly from bulge stars, occurs especially in the red domain.
Essentially the same CMD features result in both colours.

Cluster fundamental parameters are derived with solar-metallicity Padova isochrones (\citealt{Girardi02})
computed with the 2MASS \jj, \hh\ and \ks\ filters\footnote{\em stev.oapd.inaf.it/$\sim$lgirardi/cgi-bin/cmd }.
The 2MASS transmission filters produced isochrones very similar to the Johnson-Kron-Cousins
(e.g. \citealt{BesBret88}) ones, with differences of at most 0.01 in \jh\ (\citealt{TheoretIsoc}).
Considering uncertainties in the isochrone fit, we obtain an age of $1.0\pm0.1$\,Gyr, reddening
$\ejh=0.80\pm0.03$, which converts to $\ebv=2.56\pm0.12$, and $A_V=7.9\pm0.4$ (\citealt{DSB2002}). The
distance of FSR\,1744 from the Sun is $\ds=3.5\pm0.1$\,kpc. With the recently derived value of the Sun's
distance to the Galactic center $\Rgc=7.2$\,kpc (\cite{GCProp}), we conclude that FSR\,1744 is located at
a Galactocentric distance $\dgc=4.0\pm0.1$\,kpc.

\subsubsection{FSR\,89}
\label{FSR89}

\begin{figure}
\resizebox{\hsize}{!}{\includegraphics{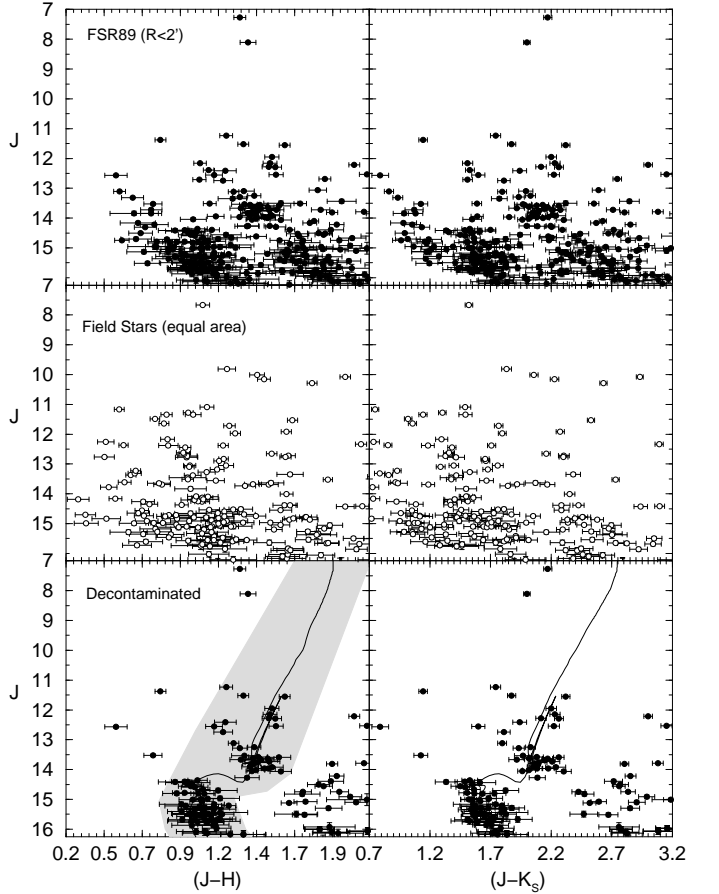}}
\caption{Same as Fig.~\ref{fig7} for the $R<2\arcmin$ region of FSR\,89. In this case
the giant clump and about 2 magnitudes of the MS are present in both CMDs. The stellar sequences
were fitted with the 1\,Gyr Solar metallicity Padova isochrone.}
\label{fig8}
\end{figure}

CMDs in both colours of the $R<2\arcmin$ region of FSR\,89 are given in Fig.~\ref{fig8}. They were
analysed likewise those of FSR\,1744 (Sect.~\ref{FSR1744}). For FSR\,89 we take as comparison
field the region $10<R(\arcmin)<20$. Similarly to FSR\,1744, the field star decontaminated CMDs (bottom panels)
present a conspicuous giant clump and about 2 magnitudes of the MS. They imply an age of
$1.0\pm0.1$\,Gyr, $\ejh=0.92\pm0.03$ ($\ebv=2.94\pm0.12$), and $A_V=9.1\pm0.4$. Its
distance from the Sun is $\ds=2.2\pm0.1$\,kpc, which puts it at $\dgc=5.4\pm0.1$\,kpc.

\subsubsection{FSR\,31}
\label{FSR31}

\begin{figure}
\resizebox{\hsize}{!}{\includegraphics{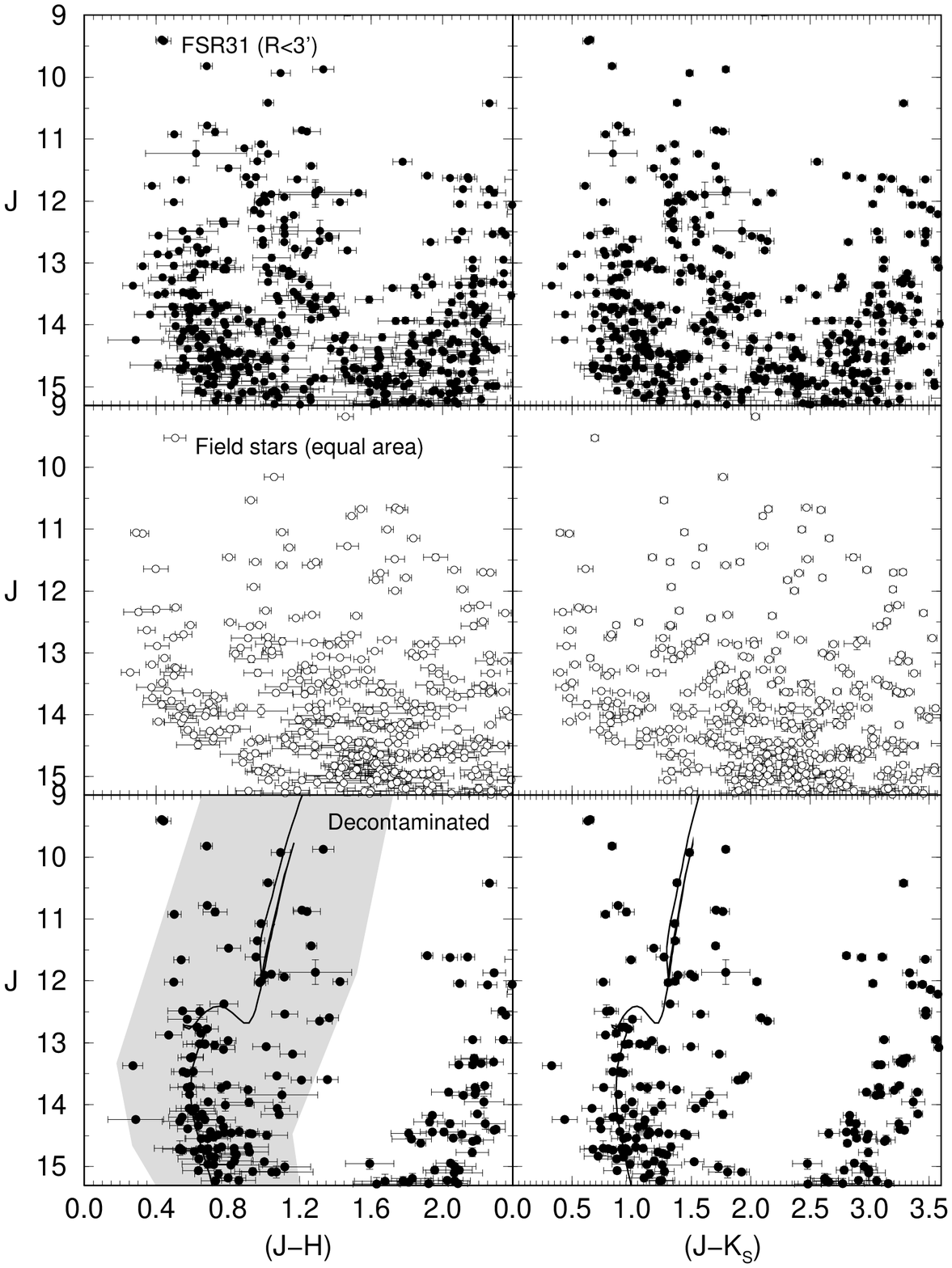}}
\caption{Same as Fig.~\ref{fig7} for the $R<3\arcmin$ region of FSR\,31. In this case
the RGB and about 3 magnitudes of the MS are present in both CMDs. The stellar sequences
were fitted with the 1.1\,Gyr Solar metallicity Padova isochrone.}
\label{fig9}
\end{figure}

Figure~\ref{fig9} shows the CMD analysis of the $R<3\arcmin$ region of FSR\,31. The comparison
field corresponds to the region $10<R(\arcmin)<50$. An interval of about 3 magnitudes of the MS, and
the RGB sequence, are present in both CMDs (bottom panels). From them we derive an age of $1.1\pm0.1$\,Gyr,
$\ejh=0.46\pm0.02$ ($\ebv=1.47\pm0.07$), and $A_V=4.6\pm0.2$. Its distance from the Sun
is $\ds=1.6\pm0.1$\,kpc, which locates it at $\dgc=5.6\pm0.1$\,kpc.

\subsubsection{Cluster ages}
\label{CluAge}

It is interesting that the present three clusters have comparable ages of $\approx1$\,Gyr.
Rather than just a coincidence, this fact probably reflects a selection effect, in the sense that
if the clusters were significantly younger with more luminous stars they would have already been
identified in previous surveys. Besides, our experience with age derivation of intermediate/old 
age OCs by means of Padova isochrones applied to field-star decontaminated CMDs shows that the 
solutions are usually rather constrained, with uncertainties smaller than 20\%\ (\citealt{BB07}). 
Conversely, this result might be providing clues on the time-scale for cluster dissolution at the 
their Galactocentric distance and respective masses.

\section{Structural parameters}
\label{Struc}

Cluster structure is investigated by means of the stellar radial density profile (RDP), built with
colour-magnitude (CM) filtered photometry (\citealt{BB07}, and references therein). CM filters are
used to remove stars with colours compatible with those of the foreground/background field. They are
wide enough to accommodate cluster MS and evolved star colour distributions, allowing for $1\sigma$
uncertainties. However, residual field stars with colours similar to those of the cluster are expected
to remain in the CM filter. This residual contamination is statistically evaluated by means of the
comparison field. CM filter widths should also account for formation or dynamical evolution-related
effects, such as enhanced fractions of binaries (and other multiple systems) towards the central parts
of clusters, since such systems tend to widen the MS (e.g. \citealt{BB07}; \citealt{N188}; \citealt{HT98};
\citealt{Kerber02}).

To avoid oversampling near the centre and undersampling at large radii, RDPs are built by counting 
stars in rings of increasing width with distance to the centre. The number and width of rings are 
adjusted to produce RDPs with adequate spatial resolution and as small as possible $1\sigma$ Poisson
errors. The residual background level of each RDP corresponds to the average number of CM-filtered
stars measured in the comparison field. The $R$ coordinate and respective uncertainty of each ring
correspond to the average position and standard deviation of the stars inside the ring.

\begin{figure}
\resizebox{\hsize}{!}{\includegraphics{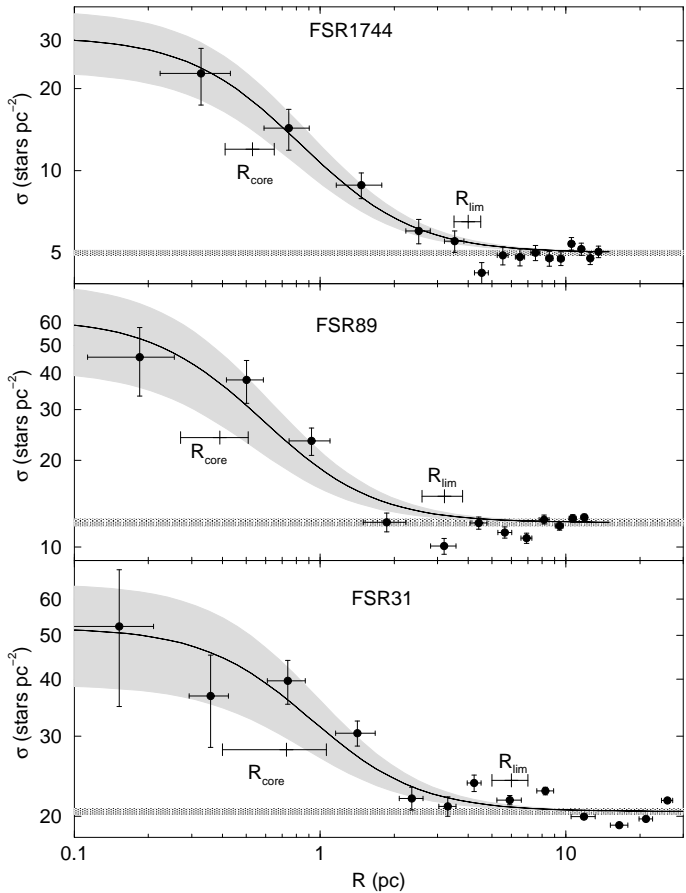}}
\caption{Stellar RDPs (filled circles) given in absolute scale. Solid lines: best-fit King-like profile.
Horizontal shaded region: stellar background level measured in the comparison field. Core and limiting
radii are indicated. Gray regions: $1\sigma$ fit uncertainty.}
\label{fig10}
\end{figure}

The resulting profiles are given in Fig.~\ref{fig10}. The RDPs are fitted with the analytical
profile $\sigma(R)=\sigma_{bg}+\sigma_{0K}/(1+(R/R_{\rm core})^2)$, where $\sigma_{bg}$ is the residual 
background density, $\sigma_{0K}$ is the central density of stars, and \rc\ is the core radius. This
function is similar to that introduced by \cite{King1962} to describe the surface brightness profiles
in the central parts of globular clusters. In all cases the adopted King-like function describes well 
the RDPs, within uncertainties. We remark that $\sigma_{0K}$ and the core radius (\rc) are derived from 
the RDP fit, while $\sigma_{bg}$ is measured in the respective comparison field. They are given in 
Table~\ref{tab4}, and the best-fit solutions are superimposed on the CM-filtered RDPs (Fig.~\ref{fig10}). 
Because of the 2MASS photometric limit, $\sigma_{0K}$ corresponds to a cutoff for stars brighter than
$\jj\approx15.6,\ \approx16$ and $\approx15.3$, respectively for FSR\,1744, FSR\,89 and FSR\,31.

\citet{Piskunov07} computed the core radii and tidal masses for a sample of 236 OCs distributed a
few kpcs from the Sun. With the core radii derived in the present paper (Table~\ref{tab4}), FSR\,31,
FSR\,1744 and especially FSR\,89, populate the small-core radius tail of the distribution given by
\citet{Piskunov07}.

The residual background contamination can be quantified by the density contrast parameter
$\delta_c=\sigma_{0K}/\sigma_{bg}$ (col.~5 of Table~\ref{tab4}). Because it is the most centrally
projected OC of the present sample, FSR\,31 presents the lowest $\delta_c$, about $1/3$ the
value of the other two OCs. Since $\delta_c$ is measured in CM-filtered RDPs, it does not necessarily 
correspond to the visual contrast produced by observed stellar distributions in 2MASS images 
(Figs.~\ref{fig3} - \ref{fig5}).

\begin{table*}
\caption[]{Structural parameters from CM-filtered photometry}
\label{tab4}
\renewcommand{\tabcolsep}{5.6mm}
\renewcommand{\arraystretch}{1.3}
\begin{tabular}{lcccccccc}
\hline\hline
&&&\multicolumn{5}{c}{RDP}\\
\cline{4-8}
Cluster&$1\arcmin$&&$\sigma_{bg}$&$\sigma_{0K}$&$\delta_c$&\rc&\rl \\
       &(pc)&&$\rm(stars\,pc^{-2})$&$\rm(stars\,pc^{-2})$&&(pc)&(pc)\\
(1)&(2)&&(3)&(4)&(5)&(6)&(7)\\
\hline
FSR\,1744&1.002&&$5.0\pm0.1$&$26\pm8$&$5.2\pm1.6$&$0.53\pm0.12$&$4.0\pm0.5$\\
FSR\,89  &0.623&&$12.2\pm0.2$&$50\pm21$&$4.1\pm1.7$&$0.39\pm0.12$ &$3.2\pm0.6$\\
FSR\,31  &0.466&&$20.5\pm0.2$&$32\pm13$&$1.5\pm0.6$&$0.73\pm0.33$ &$6.0\pm1.0$\\
\hline
\end{tabular}
\begin{list}{Table Notes.}
\item Col.~2: arcmin to parsec scale. To minimize degrees of freedom in RDP fits with the King-like 
profile (see text), $\sigma_{bg}$ was kept fixed (measured in the respective comparison fields) while 
$\sigma_{0K}$ and \rc\ were allowed to vary. Col.~5: cluster/background density contrast
($\delta_c=\sigma_{0K}/\sigma_{bg}$), measured in CM-filtered RDPs.
\end{list}
\end{table*}

We also provide in col.~7 of Table~\ref{tab4} the cluster limiting radius and uncertainty, which are
estimated by comparing the RDP (taking into account fluctuations) with the background level. \rl\ 
corresponds to the distance from the cluster centre where RDP and background become statistically
indistinguishable. For practical purposes, most of the cluster stars are contained within $\rl$. The
limiting radius should not be mistaken for the tidal radius; the latter values are usually derived from 
King (or other analytical functions) fits to RDPs, which depend on wide surrounding fields and as small 
as possible Poisson errors (e.g. \citealt{BB07}). In contrast, \rl\ comes from a visual comparison
of the RDP and background level.

We remark that the empirical determination of a cluster limiting radius depends on the RDP and 
background levels (and respective fluctuations). Thus, dynamical evolution may indirectly affect the
(measurement of the) limiting radius. Since mass segregation drives preferentially low-mass stars to the
outer parts of clusters, the cluster/background contrast in these regions tends to lower as clusters age.
As an observational consequence, smaller values of limiting radii should be measured, especially for
clusters in dense fields. However, simulations of King-like OCs (\citealt{BB07}) show that, provided not
exceedingly high, background levels may produce limiting radii underestimated by about 10--20\%. The core 
radius, on the other hand, is almost insensitive to background levels (\citealt{BB07}). This occurs because
\rc\ results from fitting the King-like profile to a distribution of RDP points, which minimizes background 
effects.

\subsection{Cluster mass}
\label{mass}

Because of the distance from the Sun and the 2MASS photometric limits (Sect.~\ref{2mass}), the MS mass
range results too narrow to allow computation of the mass function (MF) for FSR\,1744 and FSR\,89. In
those cases we simply computed the mass of the observed (background-subtracted) MS and evolved stars.
The results are $M_{obs}=160\pm33$\,\ms\ for FSR\,1744 and $M_{obs}=143\pm20$\,\ms\ for FSR\,89. For
FSR\,31 the observed mass is $M_{obs}=483\pm55$\,\ms. The MF $\left(\phi(m)\propto m^{-(1+\chi)}\right)$
of FSR\,31 could be computed, resulting in a slope $\chi=1.8\pm0.9$. Extrapolation of the MF to the
H-burning mass limit ($m=0.08\,\ms$) assuming the universal Initial Mass Function of 
\citet{Kroupa2001} produces a total mass of $M_{total}=(5.1\pm2.3)\times10^3$\,\ms. The latter value
should be taken as an upper limit, since $\sim1$\,Gyr of dynamical evolution may have driven a significant 
fraction of the low-mass content to the field (e.g. \citealt{BB07}; \citealt{DetAnalOCs}). Low-mass
stars spend most of their lives in the outer parts of clusters because their binding energies are
smaller and thus have higher probability of escape (e.g. \citealt{Aarseth71}; \citealt{Terlevich87}). 
Low-mass star depletion has been investigated in N-body simulations of star clusters in external tidal 
fields by e.g. \citet{BM03}, who found that the depletion may be strong enough to change an initially 
increasing mass function into a decreasing one towards low-mass stars. The observed mass values of FSR\,31, 
FSR\,1744 and FSR\,89 ($\sim140 - 480\,\ms$) are slightly lower than the average cluster mass of 
the tidal mass distribution given by \citet{Piskunov07}.

\section{Discussion}
\label{discuss}

The preceding sections indicate that we are dealing with Gyr-class OCs located in the inner Galaxy.
Clusters in that region are subject to important tidal interactions in the form of shocks due to disk
and bulge crossings, as well as encounters with massive molecular clouds. Over large time periods, 
these processes tend to dynamically heat a star cluster, which enhances the rate of low-mass star 
evaporation and produces an increase of the cluster in all scales. For some clusters, on the other
hand, mass segregation and evaporation may also lead to a phase of core contraction (Sect.~\ref{DynEvol}).
Consequently, these effects tend to disrupt most clusters, especially the less populous ones.

\begin{figure}
\resizebox{\hsize}{!}{\includegraphics{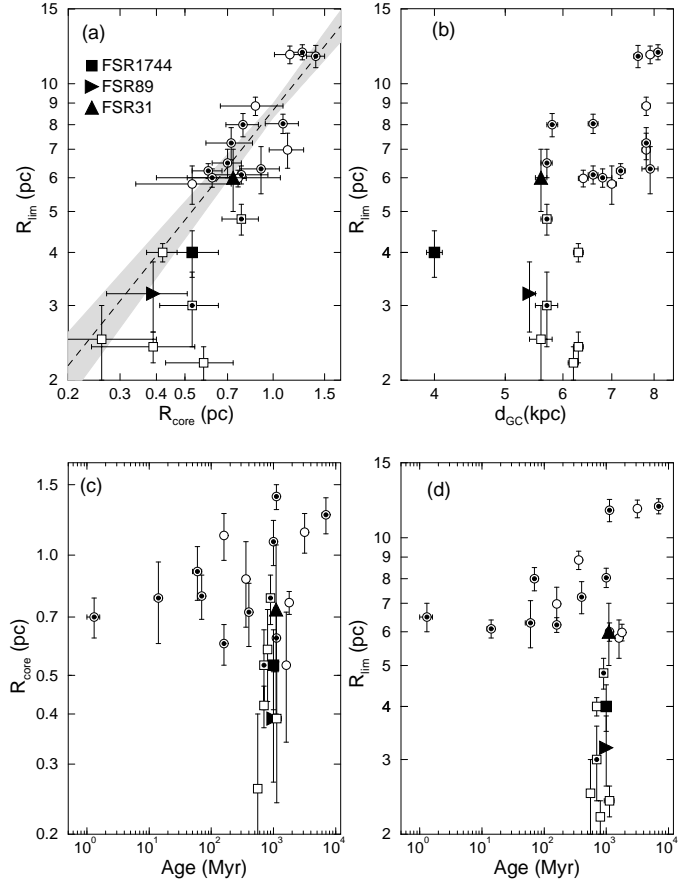}}
\caption[]{Relations involving structural parameters of OCs. Circles: nearby OCs, including
two young ones. Squares: OCs projected on dense fields towards the centre. Symbols with dots:
OCs more massive than 1\,000\,\ms.}
\label{fig11}
\end{figure}

To put FSR\,1744, FSR\,89 and FSR\,31 into perspective we compare in Fig.~\ref{fig11} their
properties with those of a sample of {\em (i)} bright nearby OCs (\citealt{DetAnalOCs}), {\em (ii)} 
young OCs (\citealt{N6611} and \citealt{N4755}), and {\em (iii)} OCs projected against the central
parts of the Galaxy (\citealt{BB07}). OCs in sample {\em (i)} have ages in the range $\rm70\la
age(Myr)\la7\,000$, masses within $\rm400\la M(\ms)\la5\,300$, and Galactocentric distances in the 
range $\rm5.8\la\dgc(kpc)\la8.1$. The OCs in sample {\em (ii)} are NGC\,6611 with 
$\rm age\approx1.3$\,Myr, $\rm M=1\,600$\,\ms\ and $\dgc=5.5$\,kpc, and NGC\,4755 with 
$\rm age\approx14$\,Myr, $\rm M=1\,150$\,\ms\ and $\dgc=6.4$\,kpc. Sample {\em (iii)} OCs are 
characterized by $\rm600\la age(Myr)\la1\,300$, $\rm210\la M(\ms)\la3\,100$, and 
$\rm5.6\la\dgc(kpc)\la6.3$.

Core and limiting radii of the OCs in samples {\em (i)} and {\em (ii)} are almost linearly related by 
$\rl=(8.9\pm0.3)\times R_{\rm core}^{(1.0\pm0.1)}$ (panel (a)), which suggests that both kinds of 
radii undergo a similar scaling, in the sense that on average, bigger clusters tend to have bigger 
cores, at least for $0.5\la\rc(pc)\la1.5$ and $5\la\rl(pc)\la15$. Linear relations between the 
core and limiting radii in OC samples were also found by \citet{Nilakshi02}, \citet{Sharma06}, and
\citet{MacNie07}. However, two-thirds of the OCs in sample {\em (iii)} do not follow that relation, 
which suggests that they are either intrinsically small or have been suffering important evaporation 
effects (Sect.~\ref{DynEvol}). The core and limiting radii of FSR\,1744, FSR\,89 and FSR\,31 are consistent 
with the relation at the $1\sigma$ level.

A dependence of open cluster size on Galactocentric distance is implied by panel (b), as previously 
suggested by \citet{Lynga82} and \citet{Tad2002}. In this context, the limiting radii of FSR\,1744,
FSR\,89 and FSR\,31 are consistent with their positions in the Galaxy. Considering the linear
relation between core and limiting radii (panel a), a similar conclusion applies to the core radius.
Part of this relation may be related to a primordial effect, in the sense that the higher density 
of molecular gas in the central Galactic regions may have produced clusters with smaller core radii, 
as suggested by \citet{vdBMP91} to explain the increase of globular cluster radii with Galactocentric 
distance.

In panels (c) and (d) we compare core and limiting radii with cluster age, respectively. This
relationship is intimately related to cluster survival/dissociation rates. Both kinds of radii 
present a similar dependence on age, in which part of the clusters - especially those more massive 
than 1\,000\,\ms - expand with time, while some seem to shrink. The bifurcation occurs at 
$\rm age\sim1$\,Gyr. FSR\,1744 and FSR\,89, the innermost OCs of the present work, have core
and limiting radii typical of the small OCs in the lower branch.

A similar effect was observed for the core radii of LMC and SMC star clusters (\citealt{Wilkinson03};
\citealt{MG03}). These clusters have core radii ($\rm0.5\la\rc(pc)\la8$) and mass ($\rm10^3\la M(\ms)\la10^6$)
significantly larger than the present ones. The core radii distribution of most LMC and SMC clusters 
is characterized by a trend of increasing \rc\ with age with an apparent bifurcation (core shrinkage)
at several hundred Myr. \citet{MG03} argument that this relationship represents true physical evolution,
with some clusters developing expanded cores due to an as yet unidentified physical process. 

At least two effects may combine to produce the smaller core radius distribution presented
by the Galactic OCs as compared to the LMC/SMC ones. The shorter core collapse time-scale
associated with lower mass clusters makes core collapse by dynamical friction a very plausible 
explanation. At the same time, probably because of their significantly lower masses, Galactic OCs 
seem to suffer more severe tidally-induced shrinkage effects than LMC/SMC clusters.

\begin{figure}
\resizebox{\hsize}{!}{\includegraphics{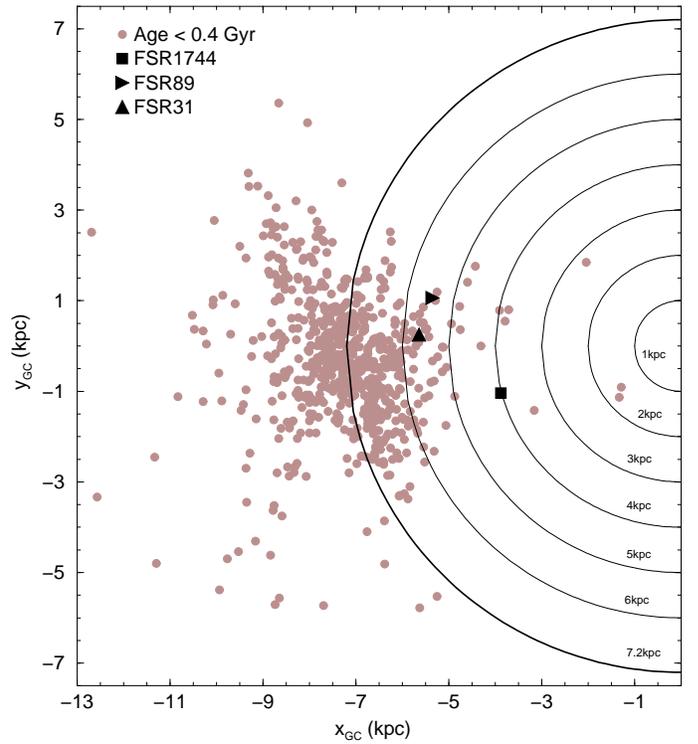}}
\caption{Same as Fig.~\ref{fig1} for the WEBDA OCs with ages $<0.4$\,Gyr (gray circles). The
present Gyr-class OCs are also shown.}
\label{fig12}
\end{figure}

\begin{figure}
\resizebox{\hsize}{!}{\includegraphics{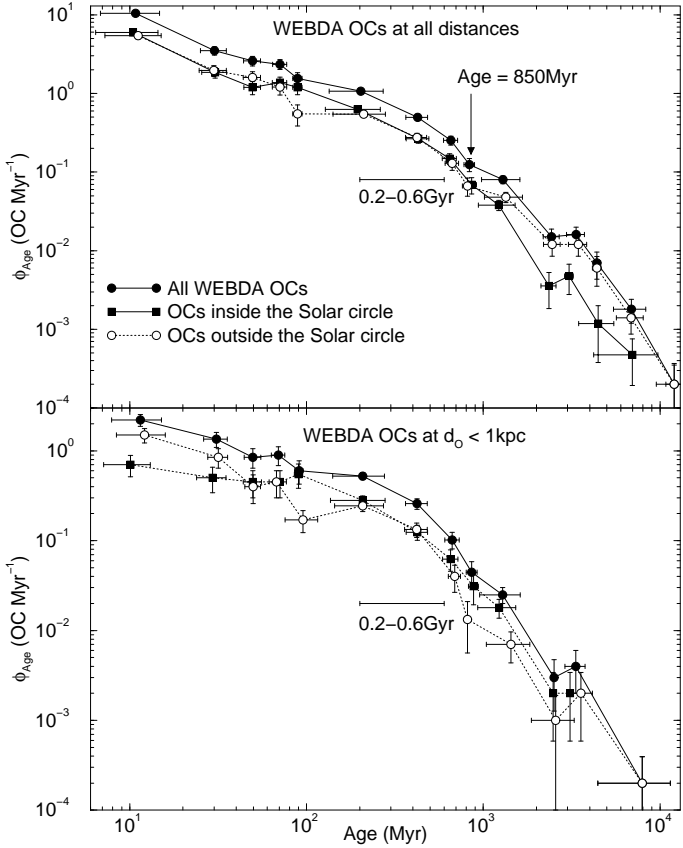}}
\caption{Top panel: age distribution function for the WEBDA OCs, including FSR\,1744, FSR\,89 and FSR\,31,
located inside (filled squares) and outside (empty circles) the Solar circle, compared to the 
combined total WEBDA sample $+$ FSR distribution (filled circles). The inside distribution 
has been normalized to the same number of OCs as outside the Solar circle. The difference in the 
OC number-density between both regions increases for ages $>850$\,Myr. The excess in the OC number 
density at age $\sim200 - 600$\,Myr is indicated. Bottom panel: same as above for a selection of OCs
with distance from the Sun $<1$\,kpc.}
\label{fig13}
\end{figure}

Besides dynamical interactions, star clusters projected against the central parts of the Galaxy -
especially the poorly-populated ones - may suffer from low-contrast effects. Such clusters
are expected to be the majority in the Galaxy (e.g. \citealt{DiskProp}). When
centrally projected, the limiting radii of poorly-populated OCs may be underestimated by about
$10 - 20\%$; the core radii, on the other hand, are not affected (\citealt{BB07}). In this sense,
the small sizes of FSR\,1744 and FSR\,89 (and to a lesser extent FSR\,31), especially the core radii,
appear to be related to dynamical effects.

Irrespective of age, inner OCs and central massive clusters are rarely detected. This is further
made clear in Fig.~\ref{fig12} where the WEBDA OCs younger than 0.4\,Gyr are plotted. On the other
hand, more than 102 embedded clusters and candidates with $|l|<30^\circ$ are catalogued. Before the
year 2003, 34 such objects were catalogued by \citet{BDB03}. In that same year, 68 new objects were
found (\citealt{REF010}; \citealt{DBSB03}) in directions of nebulae. If the number of surviving
open clusters in each generation corresponds to a fraction of $\sim4\%$ of the embedded ones
(\citealt{LL2003}), it should be expected that about 1\,000 OCs per Gyr would survive the infant 
mortality phase. However, the number of OCs of any age in the inner Galaxy is very small 
(Figs.~\ref{fig1} and Fig.~\ref{fig12}). This reminds the recurrent question whether inner 
Galaxy clusters cannot be observed because of strong absorption and crowding, or have been 
systematically dissolved by the different tidal effects (Sect.~\ref{intro}).

A more comprehensive picture of the cluster age and Galactocentric distance dependence is provided
by the OC age distribution function $\phi(\tau)=\frac{dN_{OC}}{d\tau}$. Before dealing with $\phi(\tau)$
we must consider how completeness affects the WEBDA sample. This issue has been investigated by
\citet{DiskProp} by means of simulations involving the actual 2MASS stellar background in the Galaxy. The
main results are: {\em (i)} the completeness-corrected radial distribution of OCs follows the expected
exponential-decay (disk) profile; {\em (ii)} OCs located at $\ds\la1.3$\,kpc are not much affected by
completeness; {\em (iii)} as expected, completeness is more critical for OCs inside the Solar circle;
{\em (iv)} disruption effects appear to become observationally relevant for OCs with $\ds\ga1.4$\,kpc
towards the Galactic center; and {\em (v)} OCs younger than 1\,Gyr are more affected by completeness
than the older ones. The latter occurs because young OCs are found preferentially at lower Galactic
latitudes (thus, high field-star density) than the old ones (\citealt{DiskProp}). Obviously, very young
OCs are easier to observe even in low-disk directions because of the presence of luminous stars.

As of may 2007, WEBDA contains 441 OCs with known age and distance located inside the Solar circle and
531 OCs outside it. With the inclusion also of FSR\,1744, FSR\,89 and FSR\,31, the present sample represents
an increase of $\approx50\%$ with respect to that used in \citet{DiskProp}. In Fig.~\ref{fig13} (top panel)
we show the age-distribution functions for the full sample and for the OCs inside and outside the Solar
circle. We remark that all the OCs contained in the age-distribution functions have survived the infant 
mortality phase (Sect.~\ref{DynEvol}). For a better comparison, the age-distribution function of OCs inside 
the Solar circle has been normalized to the number of OCs outside the Solar circle (531). The main 
feature of both age-distributions is that they fall off with increasing slopes with cluster age. Both are 
similarly flat, at $1\sigma$, for ages $\la850$\,Myr, which is consistent with the fact that most of these 
OCs are not old enough to have suffered significant disruption effects. Indeed, this age threshold is similar 
to the upper limit for the disruption time-scale of OCs less massive than 5\,000\,\ms, $\tdis\la900$\,Myr
(e.g. \citealt{Lamers05}). For older ages, on the other hand, the number-density of OCs inside the Solar 
circle systematically falls off with age significantly faster than outside the Solar circle. 
According to the items {\em (iii)} to {\em (v)} above, the difference in the number-density of old OCs may
be partly accounted for by completeness and partly by disruption, since $\approx42\%$ of the OCs inside the
Solar circle are more distant than $\ds=1.4$\,kpc from the Sun. In addition, contrary to outside the Solar
circle, the inner Galaxy appears not to host OCs older than $\approx7$\,Gyr. These arguments are consistent
with a Galactocentric-distance dependent OC disruption time-scale, coupled to decreasing completeness
towards the central parts.

To minimize completeness and disruption effects, we also show in Fig.~\ref{fig13} (bottom panel) the
age-distribution functions restricted to the OCs with distance from the Sun $\ds<1$\,kpc. The restricted 
age-functions contain 187 OCs inside the Solar circle and 173 outside. Within uncertainties, the distributions 
for OCs older than $\sim50$\,Myr are similar, except for the lack of OCs older than $\approx7$\,Gyr inside the 
Solar circle. Another difference is the low-frequency of OCs younger than $\sim50$\,Myr inside the 
Solar circle.

Qualitatively, the age-distribution functions for the WEBDA $+$ present FSR sample (top panel) and
especially for the $\ds<1$\,kpc sample (bottom panel) are similar to that of a restricted region,
$\ds<600$\,pc (\citealt{Lamers05}), including the excess on the OC number density for ages in the
range $\sim200 - 600$\,Myr. The shape of the age-distribution function of the OCs in this
restricted region was shown to be consistent with the combined effects of the tidal field strength,
GMC density and the dependence of disruption on cluster mass (\citealt{LG06}).

Automated infrared surveys such as that of \citet{FSR07} could provide observational clues to settle the
tidal dissolution {\em vs} completeness issue. Based on over-density arguments, they found 89 candidates 
within $|l|<30^\circ$. However, they concluded that $\approx50\%$ of their objects are possibly field 
fluctuations. Studies like Kharchenko's (2005) that use proper-motion data to sieve cluster
members, or the approach that we use which is based on CMD and radial density profile decontamination
can establish an open cluster nature. Among the 89 \citet{FSR07} candidates towards the central parts,
FSR\,1744, FSR\,89 and FSR\,31 are the best cases. A preliminary conclusion based on image inspection
is that we do not expect to find a significantly larger number of OCs in \citet{FSR07} central sample.
More studies like the present one providing decontaminated CMDs and profile information are necessary
to better determine the fraction of actual clusters. 

\section{Concluding remarks}
\label{Conclu}

Fundamental and structural parameters of the open clusters FSR\,1744, FSR\,89 and FSR\,31 were
derived in the present work. These objects are included in the catalogue of candidate star clusters
of \citet{FSR07}, which is an automated survey for overdensities in the near-infrared. The present
results are based on field-star decontaminated 2MASS CMDs and stellar density profiles, using 
analysis algorithms from \citet{BB07}, and references therein. They are confirmed to be Gyr-class 
open clusters located in the inner Galaxy, at $\dgc=4.0 - 5.6$\,kpc, with absorptions in the range
$A_V=4.6 - 9.1$\,mag. Compared to nearby open clusters, they show small core and limiting radii, 
which appears to be related to a $\sim$Gyr-long period of tidal interactions with the bulge, disk 
and possibly molecular clouds. They are apparently dynamical survivors in a region where most open
clusters are short-lived. In fact, the age-distribution of open clusters inside the Solar circle 
presents a deficiency of clusters older than $\approx850$\,Myr with respect to the outer age-distribution.

A preliminary inspection of the candidates in \citet{FSR07} for $|l|<30^\circ$ suggests that the
number of unknown star clusters must not be large, since they expect an important fraction of their
targets to be field fluctuations. However, to establish their nature the candidates should be explored 
in more detail with field-decontamination filters such as proper motion, CMD and stellar density 
profiles. Besides a natural improvement on the statistics of the open cluster parameter space,
another important result would be a better definition of the open cluster age-distribution function,
especially in the inner Galaxy. The present work represents a step in that direction.

\begin{acknowledgements}
We thank the anonymous referee for comments and suggestions.
We acknowledge partial financial support from the Brazilian agency CNPq.
\end{acknowledgements}


\end{document}